\documentclass[preprint,nofootinbib,amsmath,amssymb,aps,prd]{revtex4}

\usepackage{braket}             
\usepackage{graphicx}           
\usepackage{bm}             
\usepackage[usenames, dvipsnames]{color}
\usepackage{hyperref}           
\usepackage{mathtools}
\usepackage[normalem]{ulem} 
\usepackage[caption=false]{subfig}
\usepackage{float}
\usepackage{graphicx}

\usepackage{soul}              
\DeclareMathOperator\erf{erf}           
\DeclareMathOperator\erfc{erfc}     

\usepackage{colonequals} 

\usepackage[usenames, dvipsnames]{xcolor}

\definecolor{capri}{rgb}{0.0, 0.75, 1.0}

\begin{document}

\title{Does acceleration  assist  entanglement harvesting?}
\author{Zhihong Liu$^{1}$, Jialin Zhang$^{1}$~\footnote{Corresponding author at jialinzhang@hunnu.edu.cn}, Robert B. Mann$^{2,3}$~\footnote{rbmann@uwaterloo.ca} and Hongwei Yu$^1$~\footnote{Corresponding author at hwyu@hunnu.edu.cn}}
\affiliation{$^1$ Department of Physics and Synergetic Innovation Center for Quantum Effects and Applications, Hunan Normal University, 36 Lushan Rd., Changsha, Hunan 410081, China\\
$^2$ Department of Physics and Astronomy, University of Waterloo, Waterloo, Ontario,  Canada, N2L 3G1 \\
$^3$ Perimeter Institute for Theoretical Physics, 31 Caroline St. N., Waterloo, Ontario, Canada, N2L 2Y5}
\date{\today}
\begin{abstract}
We explore  whether acceleration assists  entanglement harvesting  for a pair
of uniformly accelerated detectors in three different  acceleration scenarios, i.e., parallel, anti-parallel
and mutually perpendicular acceleration, both in the sense of the  entanglement harvested and harvesting-achievable separation between the two detectors.
Within the framework of entanglement harvesting protocols and the Unruh-DeWitt model of detectors locally interacting with massless
scalar fields via a Gaussian  switching function with an interaction duration parameter, we find that, in the sense of  the entanglement  harvested,  acceleration is a mixed blessing insofar as it  increases the harvested entanglement for a large  detector energy gap relative to the interaction duration parameter, whilst inhibiting  the entanglement harvested for a small energy gap.  Regarding the harvesting-achievable separation range between the detectors, we further find  that for very small acceleration and large energy gap, both relative to the duration parameter,  acceleration-assisted enhancement can happen in all  three acceleration scenarios. This is in sharp contrast to what was argued previously: that the harvesting-achievable range can be enhanced only for anti-parallel acceleration. However, for a not too small acceleration relative to the duration parameter and an energy gap larger than the acceleration, we find that only detectors in parallel acceleration possess a harvesting-achievable range  larger than those at rest.
\end{abstract}

\maketitle


\section{Introduction}

The Unruh effect is a  conceptually remarkable result in quantum field
theory. It attests that observers under uniform acceleration in
the Minkowski vacuum   detect a thermal bath of particles at a
temperature proportional to the acceleration~\cite{Unruh:1976}.
Theoretically, the Unruh effect (for a review, see, for example, Ref.~~\cite{Crispino:2008}) is usually considered as the  close
``cousin" of both the Hawking  and Hawking-Gibbons effects due to its intrinsic relationship to the thermal emission of particles from black holes~\cite{Hawking:1975} and cosmological
horizons~\cite{Hawking:1977}.  Up to now, a lot of effort has been
made to understand  physical phenomena associated with  accelerated observers,  such  as the geometric
phase~\cite{EDU:2011,Hu:2012,Zhjl:2016}, the Lamb
shift~\cite{Audretsch :1995,Passante:1998,Zhu:2010} and quantum
entanglement~\cite{Benatti:2004,Fuentes-Schuller:2004iaz, Alsing:2006cj,Zhjl:2007,Landulfo:2009,Doukas:2010,Ostapchuk:2012,Hu:2015,Cheng:2018,Koga:2019,She:2019,Zhjl:2020,Majhi:2021}.

Some time ago, inspired by  the pioneering  work of Summers and
Werner~\cite{Summers:1987}  that the vacuum state of a free quantum
field can maximally violate Bell's inequalities within the framework
of the formal algebraic quantum field theory, it has been demonstrated that vacuum entanglement can  be extracted
by detectors/atoms interacting locally  with vacuum fields for a
finite time~\cite{VAN:1991,Reznik:2003}.  Much more recently this phenomenon was operationalized via the Unruh-DeWitt (UDW) detector model, and  has since
been known as entanglement
harvesting~\cite{Salton-Man:2015,Pozas-Kerstjens:2015}.  This model regards a detector as an idealized atom (or qubit) with two energy levels (ground and excited) separated by an energy gap $\Omega$. Harvesting with this model   has been
examined in a wide variety of scenarios, where it has been shown to be sensitive to
the intricate motion of detectors~\cite{Salton-Man:2015,Zhjl:2020}, the presence of
boundaries~\cite{CW:2019,CW:2020,Zhjl:2021}, and the properties of
spacetime including its dimension~\cite{Pozas-Kerstjens:2015},
topology~\cite{EDU:2016-1}, curvature
~\cite{Steeg:2009,Nambu:2013,Kukita:2017,Zhjl:2019,Robbins:2020jca,Ng:2018-2,Zhjl:2018}, and causal structure
\cite{Henderson:2020zax,Tjoa:2020,Gallock-Yoshimura:2021}.

One of the first investigations considered an interesting issue as to whether or not acceleration can
assist  entanglement harvesting~\cite{Salton-Man:2015}. Specifically rangefinding  -- the harvesting-achievable range
for the separation between detectors -- was studied for detectors accelerating in the parallel and
anti-parallel acceleration scenarios. It was
argued that   entanglement harvesting   can be enhanced only in the  anti-parallel scenario.

In the present paper, we reconsider this issue, performing a complete investigation of the acceleration-assisted entanglement harvesting phenomenon in three scenarios: parallel,  anti-parallel, and perpendicular detector accelerations. This latter case was  not considered in the original study~\cite{Salton-Man:2015}.  We emphasize that  a panoramic understanding of    acceleration-assisted entanglement harvesting should include the following two aspects.  One is whether  entanglement
harvesting can indeed be assisted by acceleration in terms of the amount of  entanglement harvested; the other is whether or not  the harvesting-achievable separation range between the detectors can be enlarged  in comparison with the inertial case?  Only the latter was analyzed before~\cite{Salton-Man:2015}.

Within the framework of the entanglement harvesting protocols, we take both aspects  into account.  We shall
consider a UDW detector  interacting locally with  quantum scalar fields via a Gaussian switching function governed by an interaction duration parameter.  All   relevant physical quantities in general can be rescaled with this parameter to be unitless.
 As we will demonstrate, acceleration is actually a mixed blessing for entanglement harvesting: it  may assist  harvesting in some circumstances and hinder it in  others. Specifically, we find for small detector gaps (relative to the duration parameter)
it generally suppresses   harvesting, whereas for large gaps it enhances  the amount of harvested entanglement. Acceleration can  either shorten or enlarge the harvesting-achievable range, depending on both the energy gap and the magnitude of the acceleration.

Our results stand in sharp contrast to what was previously argued for parallel acceleration~\cite{Salton-Man:2015}. We find that the acceleration-assisted enhancement can take place not only  in the anti-parallel acceleration scenario  but  in the other two  acceleration scenarios as well, provided  the detectors have a very small acceleration and a large energy gap. We shall see that this discrepancy appears due to an inappropriate symmetrization of the Wightman function and the use of the saddle point approximation, which may  not be appropriate in certain circumstances.

Entanglement harvesting by uniformly accelerated detectors in the presence of a
reflecting boundary was recently studied~\cite{Zhjl:2021},  the focus being on the influence of the presence of a boundary, with the  energy gap  fixed to be small (relative to the duration parameter) for simplicity. Detectors at rest were found to be  likely to harvest more entanglement than those experiencing acceleration in all scenarios within a certain fixed distance from the boundary. Furthermore, the inertial case also had a comparatively larger harvesting-achievable separation range. Namely, there was  no acceleration-assisted entanglement harvesting   for detectors located sufficiently far  from the boundary~\cite{Zhjl:2021}.  Similar results for circularly accelerated detectors --  that  harvested entanglement always degrades with increasing acceleration -- were likewise found recently~\cite{Zhjl:2020}.  This is in accord with expected intuition, since the thermal noise due to the Unruh effect  generally drives  accelerated detectors to decohere. Surprisingly, as we will demonstrate later in the present paper,
the conclusion that acceleration hinders rather than assists entanglement harvesting is however valid only when the energy gap is small relative to the interaction duration parameter. In fact, if we relax the restriction of a small energy gap, then the conclusions  significantly change, and the phenomenon of acceleration-assisted entanglement harvesting does occur.

The rest of our paper is organized as follows. In the next section, we review the basics for   UDW detectors locally interacting with vacuum scalar fields, and give the trajectories of uniformly accelerated detectors in the parallel, anti-parallel, and perpendicular acceleration scenarios as well as the corresponding Wightman functions. In section III, we estimate the  entanglement  harvested by numerical calculations,  probing  whether the acceleration could assist the entanglement harvesting both in the sense of  the entanglement harvested  and harvesting-achievable separation. Finally, we conclude with a summary in section IV.   For convenience,  we adopt the natural units $\hbar=c=1$ throughout this paper.


\section{Setup}
In this section, we first review the general model of accelerated
detectors in local interaction with quantum fields  in  Minkowski spacetime. The concrete
Wightman functions for accelerated detectors  in different acceleration scenarios are given
explicitly. In the framework of the entanglement
harvesting protocols, the general expression of the concurrence as a
measure of the entanglement is provided in detail.

\subsection{The general model}
We consider two identical  point-like detectors labeled by $A$ and $B$
with an  energy gap $\Omega$  between the ground state
$|0_{D}\rangle$ and the excited state $|1_{D}\rangle$ \textbf{($D\in\{A,B\}$)},  parameterize the classical spacetime trajectory of the detector  by its proper time $\tau$,   denote the massless scalar
field that the detectors are in interaction with by  $\phi\left[x_D(\tau)\right]$,  and assume that the  interaction Hamiltonian between the detector and
the field in the interaction picture is given by
 \begin{equation}\label{Int1}
 H_{D}(\tau)=\lambda \chi(\tau)\left[e^{i \Omega\tau} \sigma^{+}+e^{-i \Omega\tau} \sigma^{-}\right] \phi\left[x_{D}(\tau)\right]\;,~~ D\in\{A,B\}.
 \end{equation}
Here, $\lambda$ is the coupling strength,
$\sigma^{+}=|1_{D}\rangle\langle0_{D}|$  and
$\sigma^{-}=|0_{D}\rangle\langle1_{D}|$ denote SU(2) ladder operators, and $\chi(\tau)=\exp [-{\tau^{2}}/(2
\sigma^{2})]$ is the Gaussian switching function  which controls the duration of interaction via the parameter $\sigma$.

Initially,   the two detectors are prepared in their ground state and the field is in a vacuum state $|0_M\rangle$.  According to the
detector-field interaction Hamiltonian~(\ref{Int1}), the final
state of the system (two detectors)  can be obtained in the basis
$\{\ket{0_A}\ket{0_B},\ket{0_A}\ket{1_B},\ket{1_A}\ket{0_B},\ket{1_A}\ket{1_B}\}$  by performing standard
perturbation theory to yield \cite{Pozas-Kerstjens:2015,Zhjl:2019}
\begin{align}\label{rhoAB}
\rho_{AB}=\begin{pmatrix}
1-P_A-P_B & 0 & 0 & X \\
0 & P_B & C & 0 \\
0 & C^* & P_A & 0 \\
X^* & 0 & 0 & 0 \\
\end{pmatrix}+{\mathcal{O}}(\lambda^4)\;,
\end{align}
where the transition probability reads
\begin{equation}\label{PAPB}
 P_D:=\lambda^{2}\iint d\tau d\tau' \chi(\tau) \chi(\tau') e^{-i \Omega(\tau-\tau')}
 W\left(x_D(t), x_D(t')\right)\quad\quad D\in\{A, B\}\;,
\end{equation}
and the quantities $C$ and $X$, which characterize nonlocal correlations, are given by\footnote{ Note that the definition of $X$ here is adopted in accordance with that in the later literature~\cite{CW:2019,CW:2020,Zhjl:2019,Zhjl:2018,Zhjl:2021} and it is just the complex conjugate of the definition of $X$ in the early Refs.~\cite{Salton-Man:2015,Steeg:2009,EDU:2016-1,Nambu:2013}.}
\begin{align}
C &:=\lambda^{2} \iint d \tau d \tau^{\prime} \chi(\tau) \chi(\tau')
e^{-i \Omega(\tau-\tau')} W\left(x_{A}(t), x_{B}(t')\right)\;,
\end{align}
\begin{align}\label{xxdef}
X:=-\lambda^{2} \iint d\tau d \tau' \chi(\tau)\chi(\tau')  e^{-i\Omega( \tau+\tau')}
\Big[\theta(t'-t)W\left(x_A(t),x_B(t')\right)+\theta(t-t')W\left(x_B(t'),x_A(t)\right)\Big]\;,
\end{align}
where  $W(x,x'):=\bra{0_M}\phi(x)\phi(x')\ket{0_M}$ is the Wightman function of the field and $\theta(t)$
represents the Heaviside theta function. Note that the detector's
coordinate time is a function of its proper time in the above equations.  
The amount of entanglement between the detectors can be quantified  by the concurrence $\mathcal{C}(\rho_{A B})$ of the final state of the two detectors. For an $X$-type density
matrix~(\ref{rhoAB}),  the concurrence takes the following simple form~\cite{EDU:2016-1,Zhjl:2018,Zhjl:2019}
\begin{equation}\label{condf}
\mathcal{C}(\rho_{A B})=2 \max \Big[0,|X|-\sqrt{P_{A}
P_{B}}\Big]+\mathcal{O}(\lambda^{4})\;.
\end{equation}

 Once the Wightman function and the particular detectors'  trajectories  are  given,  the amount  of entanglement harvested by the detectors from the fields can be obtained straightforwardly. Specifically, for two detectors at rest, the harvested entanglement (concurrence)  can be analytically computed  from~(\ref{condf}) with the transition probability $P_D$ and  the non-local correlation term $X$ given as follows~\cite{EDU:2016-1,Zhjl:2021}
\begin{align}
P_{D} =\frac{\lambda^{2}}{4 \pi}\left[e^{-\Omega^{2} \sigma^{2}}-\sqrt{\pi}
\Omega \sigma \operatorname{Erfc}(\Omega
\sigma)\right]\qquad X=\frac{-i\lambda^2}{4\sqrt{\pi}} \frac{\sigma}{L}  e^{-\sigma ^2 \Omega ^2 - L^2/(4\sigma^2)}  \erfc\left(i\frac{L}{2\sigma}\right)
\end{align}
where $\erf(x)$ is the error function and $\erfc(x):=1-\erf(x)$.
\subsection{ Acceleration scenarios  }
\begin{figure}[!htbp]
\centering
\subfloat[]{\includegraphics[width=0.3\linewidth]{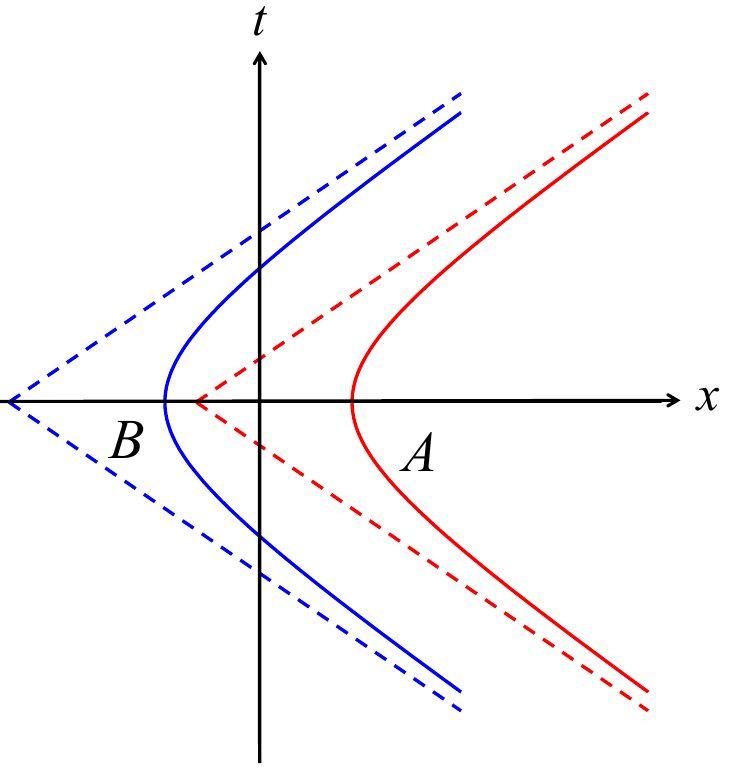}}\quad
 \subfloat[]{\includegraphics[width=0.3\linewidth]{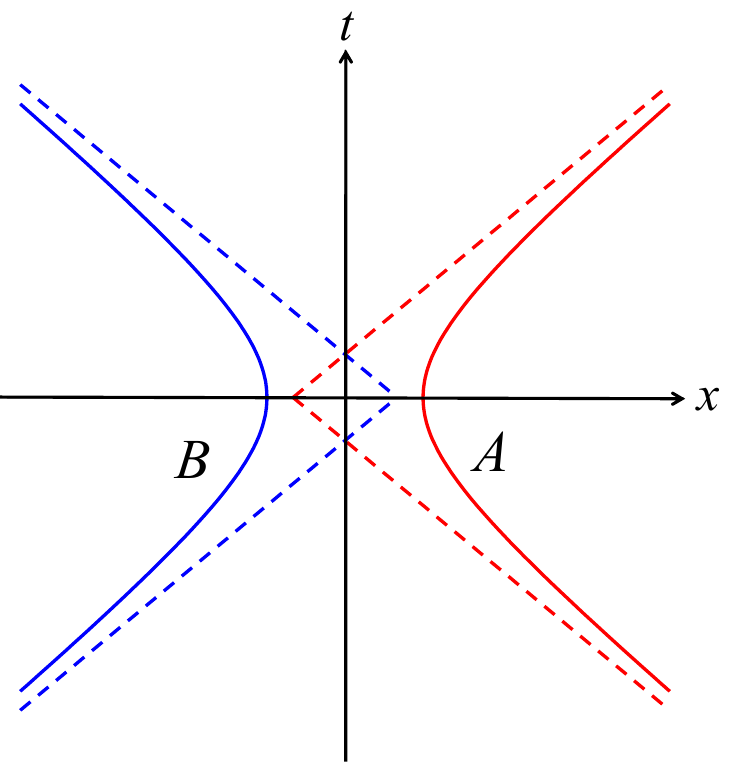}}\quad
 \subfloat[]{\includegraphics[width=0.32\linewidth]{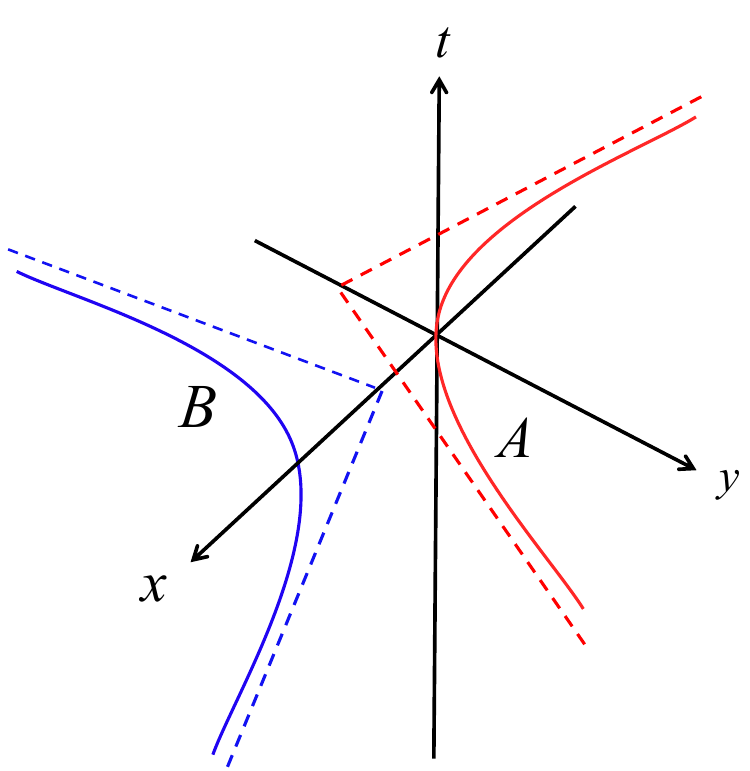}}
\caption{The worldlines of two uniformly accelerated detectors in
three acceleration scenarios:  the parallel acceleration (left),
the anti-parallel acceleration(middle) and the perpendicular
acceleration (right).}\label{model}
\end{figure}

 We now describe different setups of two detectors in acceleration.  The concrete acceleration scenarios we are going to consider  include that of the parallel, anti-parallel and mutually perpendicular acceleration (see the wordlines in Fig.(\ref{model})).

\subsubsection{Parallel acceleration}
Suppose that two detectors are accelerated along the $x$-direction with  an acceleration $a$. The trajectories can be written as~\cite{Salton-Man:2015}
\begin{align}\label{traj-ua}
&x_A:=\{t=a^{-1}\sinh(a\tau_A)\;,~x=a^{-1}[\cosh(a\tau_A)-1]+\frac{L}{2}\;,~y=z=0\}\;,
\nonumber\\
&x_B:=\{t=a^{-1}\sinh(a\tau_B)\;,~x=a^{-1}[\cosh(a\tau_B)-1]-\frac{L}{2}\;,~y=z=0\}\;,
\end{align}
where $L$ represents the separation between such detectors, as measured  by an inertial observer at a fixed $x$ (i.e., in the laboratory reference frame).

The Wightman function for massless scalar
fields in four dimensional Minkowski spacetime is given by~\cite{Birrell:1984}
\begin{align}\label{wigh-1}
W\left(x, x'\right)=-\frac{1}{4 \pi^{2}}\frac{1}{(t-t'-i
\epsilon)^{2}-|\mathbf{x}-\mathbf{x'}|^{2}}\;.
\end{align}
The transition probability for a uniformly accelerated detector  can be derived by substituting  trajectories~(\ref{traj-ua}) into~(\ref{wigh-1}) and then computing~(\ref{PAPB}).  After some algebraic  manipulations, we  arrive at~\cite{Zhjl:2020,Zhjl:2021}
\begin{align}\label{PD-1}
P_{D} =&\frac{\lambda^{2} a \sigma}{4 \pi^{3 / 2}} \int_{0}^{\infty}
d\tilde{s} \frac{\cos (\tilde{s} \beta)e^{-\tilde{s}^{2}
\alpha}\left(\sinh ^{2}\tilde{s}-\tilde{s}^{2}\right)}{\tilde{s}^{2}
\sinh^{2} \tilde{s}}+\frac{\lambda^{2}}{4 \pi}\left[e^{-\Omega^{2}
\sigma^{2}}-\sqrt{\pi} \Omega \sigma \operatorname{Erfc}(\Omega
\sigma)\right]\;,
\end{align}
where $\beta=2\Omega/ a$ and $\alpha=1 /(a \sigma)^{2}$.

Similarly, by substituting  Eqs.~(\ref{traj-ua}) and (\ref{wigh-1})
into Eq.~(\ref{xxdef}), the nonlocal correlation term $X$, denoted
here by $X_{((}$ in the parallel acceleration case, can be found to be
\begin{align}\label{XPP} X_{((}&=-\lambda^{2}\int_{-\infty}^{\infty}
d\tau\int_{-\infty}^{\tau} d \tau' \chi(\tau)\chi(\tau')
e^{-i\Omega( \tau+\tau')}
\Big[W\left( x_A(\tau'),x_B(\tau)\right)+W\left(x_B(\tau'),x_A(\tau)\right)\Big]\nonumber\\
&=- \lambda^2\int_{0}^{\infty} d\tilde{y} F(\tilde{y})\;,
\end{align}
where we have  introduced two variables $\tilde{x}=\tau+\tau'$,
$\tilde{y}=\tau-\tau'$ and an auxiliary function
\begin{equation}\label{XPP-22}
 F(\tilde{y}):=\int_{-\infty}^{\infty}d\tilde{x} e^{-\frac{\tilde{x}^2+\tilde{y}^2}{4\sigma^2}-i \tilde{x}\Omega}D_{((}(\tilde{x},\tilde{y})
\end{equation}
with the symmetrized Wightman function $D_{((}(\tilde{x},\tilde{y})$  given by\footnote{This expression disagrees with the
 Wightman function $D_{((}$ employed in Ref.~\cite{Salton-Man:2015}, specifically the 2nd line of Eq.(2.14), which included only a term equivalent to the first term in \eqref{Dparr}. This first term is not symmetric under the exchange $L\leftrightarrow -L$, and so
  does not keep $X_{((}$ invariant under the exchange $x_A\leftrightarrow{x_B}$; in order to maintain this symmetry,
both terms in   \eqref{Dparr} are required. Note in \eqref{Dparr} that an {\it overall} positive coefficient in $i\epsilon$  has
been absorbed into $\epsilon$.}
\begin{align}\label{Dparr}
D_{((}(\tilde{x},\tilde{y}):=&\frac{1}{2}\Big[W\left(
x_A(\tau'),x_B(\tau)\right)+W\left(x_B(\tau'),x_A(\tau)\right)\Big]\nonumber\\=&\frac{a^2}{32\pi^2}\Bigg\{\bigg[\Big(\frac{aL}{2}-e^{-\frac{\tilde{x}a}{2}}\sinh\frac{\tilde{y}a}{2}\Big)\Big(\frac{aL}{2}+e{^\frac{\tilde{x}a}{2}}\sinh\frac{\tilde{y}a}{2}\Big)-i\epsilon\bigg]^{-1}
\nonumber\\&+\bigg[\Big(\frac{aL}{2}+e^{-\frac{\tilde{x}a}{2}}\sinh\frac{\tilde{y}a}{2}\Big)\Big(\frac{aL}{2}-e{^\frac{\tilde{x}
a}{2}}\sinh\frac{\tilde{y}a}{2}\Big)-i\epsilon\bigg]^{-1}\Bigg\}\;.
\end{align}

In general, the double integral  for $X_{((}$ can be numerically evaluated
by first integrating
Eq.~(\ref{XPP-22}) in the sense of the Cauchy principal value  as long as the Wightman
functions are treated as the well-defined
distributions~\cite{Bogolubov:1990}. Here, it is easy to see that
$F(\tilde{y})$  is an even function of the detectors'
separation,  so that $X_{((}$ is invariant under  an exchange between
detector $A$ and $B$ as expected.

\subsubsection{Anti-parallel acceleration}
For the anti-parallel acceleration scenario, the trajectories for two
accelerated detectors can be simply written in the following
form~\cite{Salton-Man:2015}
\begin{align}\label{traj-aa}
&x_A:=\{t=a^{-1}\sinh(a\tau_A)\;,~x=a^{-1}[\cosh(a\tau_A)-1]+\frac{L}{2}\;,~y=z=0\}\;,
\nonumber\\
&x_B:=\{t=a^{-1}\sinh(a\tau_B)\;,~x=-a^{-1}[\cosh(a\tau_B)-1]-\frac{L}{2}\;,~y=z=0\}\;,
\end{align}
 where  $L$ represents  the separation between the detectors at their
closest approach (i.e., the minimum distance at the origin of the
time coordinate) as seen by a rest observer located at a constant
$x$, and $a$ again stands for the magnitude of the acceleration along the
$x$-axis. Here, it is worth pointing out that the nonzero $L$ in the
trajectories~(\ref{traj-aa}) generally relaxes the condition of
overlapping apexes shared by four Rindler wedges of two
detectors~\cite{Salton-Man:2015}. The transition probability still
has the same  form  as  Eq.~(\ref{PD-1}),  and  the nonlocal
correlation term, now denoted by $X_{)(}$ in the anti-parallel
acceleration scenario, can be straightforwardly  deduced by
substituting Eq.~(\ref{traj-aa}) into Eq.~(\ref{xxdef})
\begin{align}\label{Xanti}
X_{)(}=- \lambda^2 \int_{-\infty}^{\infty} d\tilde{x}
\int_{0}^{\infty} d\tilde{y}
e^{-\frac{\tilde{x}^2+\tilde{y}^2}{4\sigma^2}-i
\tilde{x}\Omega}{D_{)(}(\tilde{x},\tilde{y})}\;,
\end{align}
where
\begin{align}\label{D-anti1}
 D_{)(}(\tilde{x},\tilde{y}) =& \frac{a^2}{16\pi^2}  \Bigg\{\bigg[e^{-\frac{\tilde{y}a}{2}}\Big(\frac{L
a}{2}-1\Big) +\frac{ae^{-a\tilde{y}/2}}{2}i\epsilon +\cosh\frac{\tilde{x} a}{2}\bigg] \bigg[e^{\frac{\tilde{y}a}{2}}\Big(\frac{L
a}{2}-1\Big) -\frac{ae^{a\tilde{y}/2}}{2}i\epsilon +\cosh\frac{\tilde{x} a}{2}\bigg]\Bigg\}^{-1}\nonumber\\
=&\frac{a^2}{16\pi^2}\Bigg\{\bigg[e^{-\frac{\tilde{y}a}{2}}\Big(\frac{L
a}{2}-1\Big)+\cosh\frac{\tilde{x} a}{2}\bigg]
\bigg[e^{\frac{\tilde{y}a}{2}}\Big(\frac{L
a}{2}-1\Big)+\cosh\frac{\tilde{x}
a}{2}\bigg]-i\epsilon a\sinh\left(\frac{\tilde{y}a}{2}\right)\cosh\left(\frac{\tilde{x}a}{2}\right)\nonumber\\& +\frac{a^2\epsilon^2}{4}\Bigg\}^{-1}\;.
\end{align}
Absorbing the positive factor $ a\sinh\left(\frac{\tilde{y}a}{2}\right)\cosh\left(\frac{\tilde{x}a}{2}\right)$ into $\epsilon$ in the second line of Eq.~(\ref{D-anti1}) and 
 discarding the $\epsilon$-squared  term, we have\footnote{Here an error in the Wightman function~(2.15)
in Ref.~\cite{Salton-Man:2015} has been corrected. Because the positive factors $\frac{ae^{-a\tilde{y}/2}}{2}$ and $\frac{ae^{a\tilde{y}/2}}{2}$  in the  1st line of Eq.~(\ref{D-anti1}) are different,  
one cannot  respectively absorb them into $\epsilon$ at first. Otherwise, the sign of $i\epsilon$ term in the denominator of the Wightman function may be changed. Note also that here the sign of the
$i\epsilon$  term  in the denominator of the Wightman function 
is negative 
regardless of the detectors' separation, in contrast to Eq.~(2.15)
in Ref.~\cite{Salton-Man:2015}.}
\begin{equation}\label{D-anti2}
D_{)(}(\tilde{x},\tilde{y})=\frac{a^2}{16\pi^2}\Bigg\{\bigg[e^{-\frac{\tilde{y}a}{2}}\Big(\frac{L
a}{2}-1\Big)+\cosh\frac{\tilde{x} a}{2}\bigg]
\bigg[e^{\frac{\tilde{y}a}{2}}\Big(\frac{L
a}{2}-1\Big)+\cosh\frac{\tilde{x}
a}{2}\bigg]-i\epsilon\Bigg\}^{-1}\;.
\end{equation}

\subsubsection{Perpendicular accelerations}

 Regarding  the perpendicular acceleration case,  the spacetime trajectories of  detectors  can be
 simply written as
\begin{align}\label{traj-va}
&x_A:=\{t=a^{-1}\sinh(a\tau_A)\;,~y=a^{-1}[\cosh(a\tau_A)-1]\;,x=z=0\}\;,
\nonumber\\
&x_B:=\{t=a^{-1}\sinh(a\tau_B)\;,~x=a^{-1}[\cosh(a\tau_B)-1]+L\;,y=z=0\}.
\end{align}
with $a$ the acceleration  magnitude and  $L$ the detectors'
separation at the closest approach.

Similarly,  the nonlocal correlation term  in the present case
denoted by $X_{\perp}$ is given by
\begin{equation}
X_{\perp}=-{\lambda^2} \int_{-\infty}^{\infty} d\tilde{x}
\int_{0}^{\infty} d\tilde{y}
 e^{-\frac{\tilde{x}^2+\tilde{y}^2}{4\sigma^2}-i \tilde{x}\Omega} D_{\perp}(\tilde{x},\tilde{y})
\end{equation}
where
\begin{equation}
 D_{\perp}(\tilde{x},\tilde{y}) =\frac{a^2}{8\pi^2}\Big[\frac{1}{h_{+}(\tilde{x},\tilde{y})}+\frac{1}{h_{-}(\tilde{x},\tilde{y})}\Big]
\end{equation}
with
\begin{align}
h_{\pm}(\tilde{x},\tilde{y}) =&3+(a L-1)^2-4
\cosh(a\tilde{x}/2)\cosh(a\tilde{y}/2)-\cosh(a\tilde{y})\nonumber\\&+\cosh(a\tilde{x})+2a
L\cosh [a (\tilde{x}\pm{\tilde{y}})/2]-i \epsilon\;.
\end{align}

Although the basic formulae for entanglement harvesting are given above
explicitly, exact analytic  results  are not easy to obtain
due to the complicated double integrals. In what follows, we resort to numerical computation.

\section{Numerical results}
\begin{figure}[!htbp]
\centering
\subfloat[$\Omega\sigma=0.01$]{\label{comp-aL11}\includegraphics[width=0.32\linewidth]{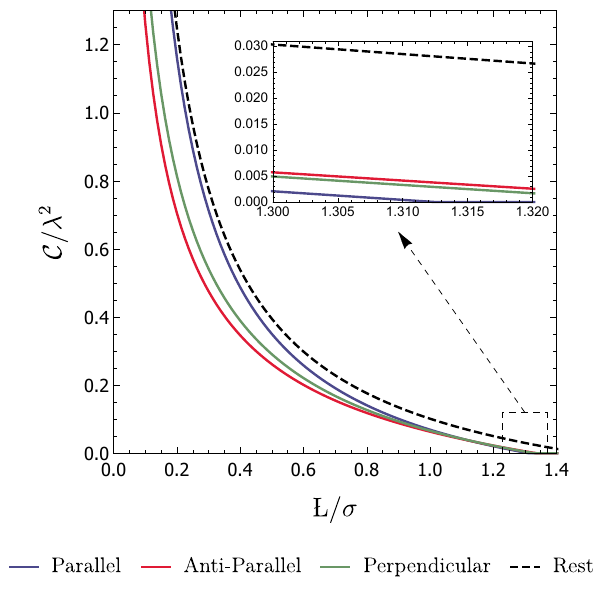}}\;
 \subfloat[$\Omega\sigma=0.50$]{\label{comp-aL22}\includegraphics[width=0.32\linewidth]{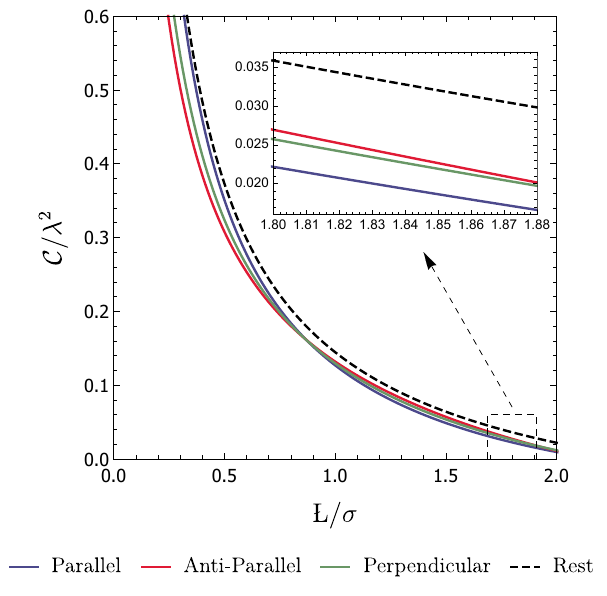}}\;
 \subfloat[$\Omega\sigma=2.00$]{\label{comp-aL33}\includegraphics[width=0.32\linewidth]{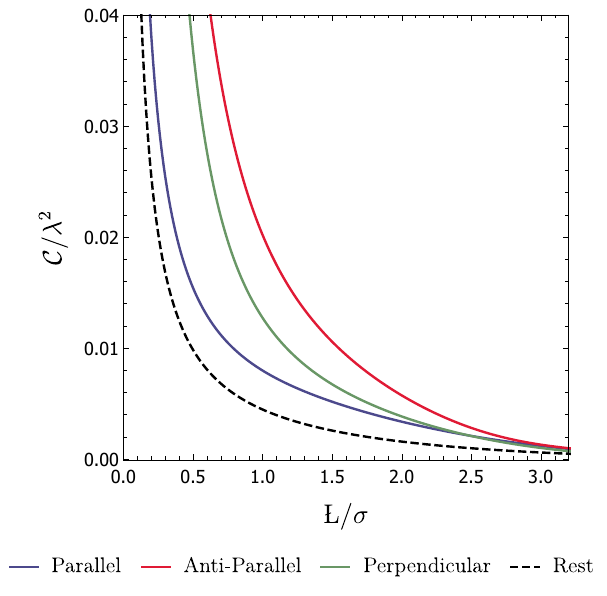}}
\caption{The concurrence is plotted as a function of $L/\sigma$  with fixed $a\sigma=0.50$, and $\Omega\sigma=\{0.01,0.50,2.00\}$ from  left to right. Here,  different colored solid lines correspond to different acceleration scenarios, and the dashed line indicates the situation of detectors at rest. For convenience, all other physical quantities are expressed in  units of  parameter $\sigma$.}\label{convsL2}
  \end{figure}

\begin{figure}[!htbp]
\centering
\subfloat[$\Omega\sigma=0.01$]{\label{comp-La11}\includegraphics[width=0.32\linewidth]{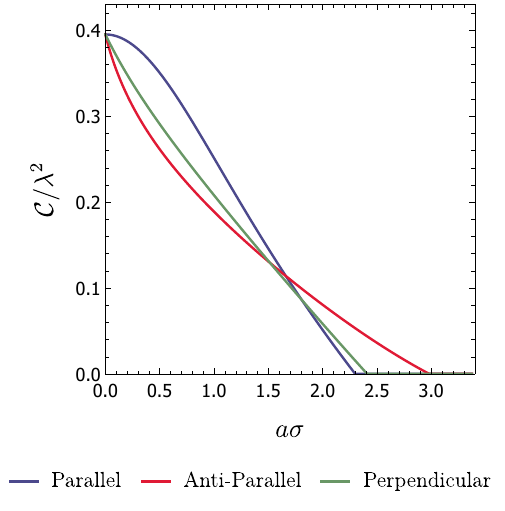}}\;\;
 \subfloat[$\Omega\sigma=0.50$]{\label{comp-La22}\includegraphics[width=0.32\linewidth]{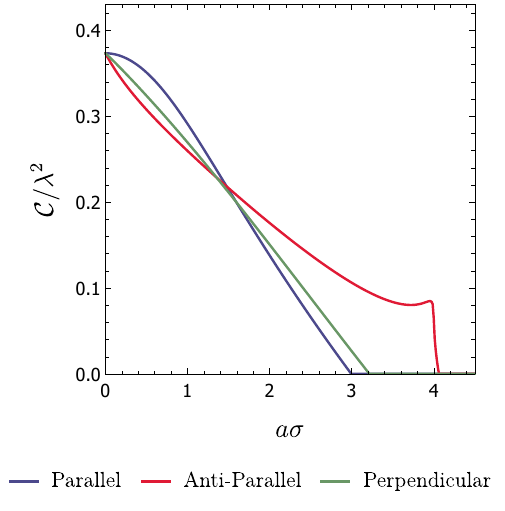}}\;\;\subfloat[$\Omega\sigma=2.00$]{\label{comp-La33}\includegraphics[width=0.32\linewidth]{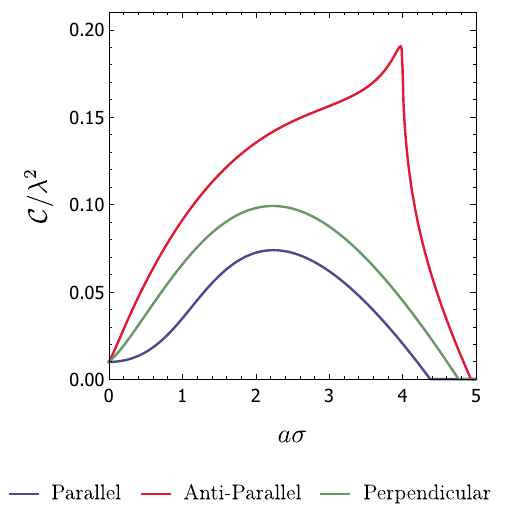}}
\caption{The concurrence, $\mathcal{C}(\rho_{A B})/\lambda^2$, is plotted as a function of
  $a\sigma$ for various energy gaps $\Omega\sigma=\{0.01,0.50,2.00\}$ with  $L/\sigma=0.50$. In the limit of $a\sigma\rightarrow0$,  the concurrence $\mathcal{C}(\rho_{A B})$ approaches the corresponding value for detectors at rest in  Minkowski spacetime. }\label{convsa}
 \end{figure}

In this section, we  will consider how
  acceleration 
 affects entanglement harvesting,
first exploring  the acceleration-assisted entanglement harvesting in
the proposed acceleration scenarios in  the sense  of the entanglement
harvested. We  begin the discussion by plotting the concurrence  as a
function of the detectors' separation $L/\sigma$ (Fig.~(\ref{convsL2})) or the acceleration
$a\sigma$  (Fig.~(\ref{convsa})) at various  energy gaps $\Omega\sigma$ for  all three scenarios.

For a small energy gap ($\Omega\sigma<1$), we find that
the harvested entanglement generally degrades with increasing
acceleration  and with increasing detector separation. As a consequence,
inertial detectors (at rest)  harvest more entanglement. In this sense,  acceleration  suppresses entanglement harvesting due to the strong thermal noise associated with the Unruh effect. Similar results have also  been found   for accelerated detectors in four-dimensional Minkowski
spacetime with a reflecting boundary~\cite{Zhjl:2021}.

Upon closer inspection, some interesting features emerge.  A cross-comparison of the concurrences
 for small accelerations, small separations, and small gaps,
 indicates that the parallel scenario harvests comparatively more entanglement than the perpendicular one, which in turn
 harvests more than the antiparallel case.  However we see that the parallel case has the most rapid decrease and so
 a crossover effect occurs for both sufficiently large separation and acceleration,
 with the rank-ordering reversed: antiparallel harvesting the most and parallel the least.  The point of crossover is gap-dependent, as is
 clear upon comparison of the left diagrams with the middle ones in Fig.(\ref{comp-aL11})\&(\ref{comp-aL22}) or
Fig.(\ref{comp-La11})\&(\ref{comp-La22})).  Similar effects
were reported in Ref.~\cite{Zhjl:2021}.

For sufficiently large gaps
($\Omega>L/\sigma^2$ and $\Omega>a$)  the thermal noise caused by  acceleration ceases to play a significant role  in  entanglement harvesting.
As shown in Fig.~(\ref{comp-La33}),   concurrence is no longer a monotonically
decreasing function of acceleration.  In fact,  for small $a\sigma$, the concurrence is  instead a  monotonically increasing function
of $a\sigma$,  meaning that the detectors in all  three acceleration scenarios
are likely to harvest more entanglement from the fields than the
detectors at rest. This is in sharp contrast to the aforementioned results for a small energy gap.   The concurrence maximizes in all scenarios at some intermediate value of the acceleration, peaking most strongly in the antiparallel scenario, with the parallel scenario harvesting the smallest amount throughout.

Notably  interesting features appear in Fig.~(\ref{comp-aL33}), which plots concurrence against separation for a large gap.  We see the
rank-ordering from smallest  to largest is parallel to perpendicular to antiparallel, as in Fig.~(\ref{comp-La33}), but with the remarkable feature that all three scenarios harvest more entanglement than the situation where the detectors are at rest.   As separation increases, concurrence decreases, with the curves tending to merge at large $L$.

\begin{figure}[!htbp]
\centering
\subfloat[$L/\sigma=0.20$]{\label{convsomega11}\includegraphics[width=0.32\linewidth]{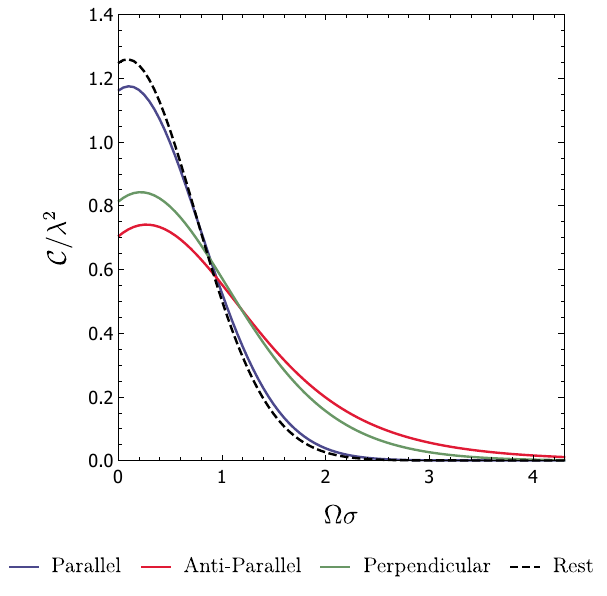}}\;
 \subfloat[$L/\sigma=0.50$]{\label{convsomega22}\includegraphics[width=0.32\linewidth]{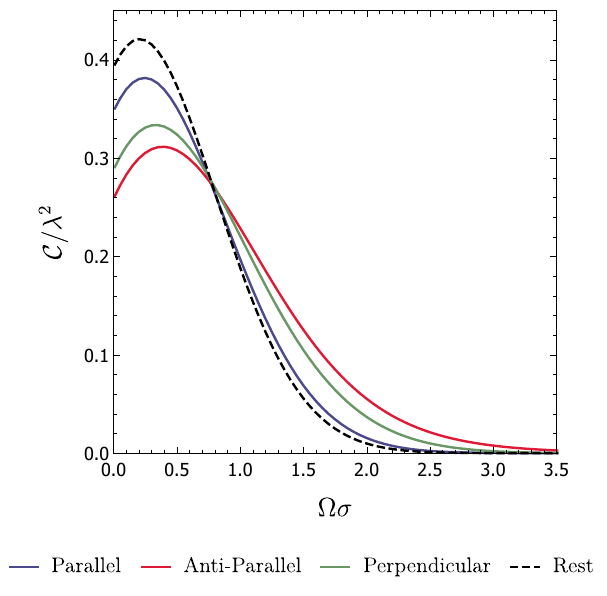}}\;\subfloat[$L/\sigma=2.00$]{\label{convsomega32}\includegraphics[width=0.33\linewidth]{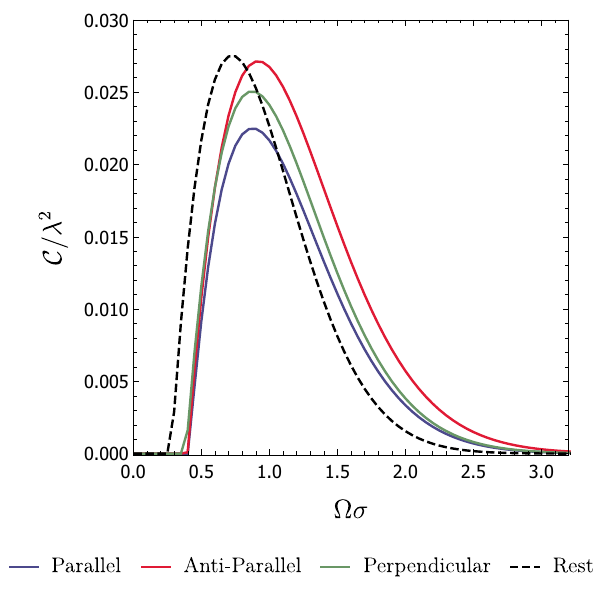}}
\caption{The concurrence is plotted as the function of $\Omega\sigma$  with $a\sigma=0.50$  and $L/\sigma=\{0.20,0.50,2.00\}$ in left-to-right order. The dashed line indicates the  detectors is at rest.}\label{convsomega1}
\end{figure}

To gain a better understanding of the influence of the energy gap, we plot concurrence  as a
function of  $\Omega\sigma$  for all acceleration scenarios in
Fig.~(\ref{convsomega1}).  For small separations $L/\sigma  \lesssim 1$, concurrence grows with increasing $\Omega\sigma$, reaching a maximum at some $\Omega\sigma < 1$ in each scenario, the ascending rank-ordering being  antiparallel to perpendicular to parallel, with the rest case harvesting the largest amount of entanglement.  However once  $\Omega\sigma > 1$ the situation dramatically changes: the rank-ordering is reversed, with inertial detectors (at rest) harvesting the least amount of entanglement.

For large separations, we see from Fig.~(\ref{convsomega32}) that the situation changes yet again.   There is a peak in the concurrence at some value of
$\Omega\sigma$ for all acceleration scenarios, with the parallel to perpendicular to antiparallel ordering from lowest to highest concurrence holding over almost the  entire gap range.  But the inertial scenario differs from these: it is larger than the acceleration scenarios for
a small $\Omega\sigma$, but peaks earlier, and is subsequently overtaken by the other cases.

Similar peaking behavior of the concurrence has also been
discovered in some spacetimes with nontrivial
topologies~\cite{Zhjl:2018,Zhjl:2019,Robbins:2020jca}. Indeed, the value of the peak
is strongly dependent on  $L/\sigma$, and the larger the
detectors' separation, the smaller the peak value.

We again find a contrast with   Ref.~\cite{Salton-Man:2015}, where it was argued that there is
an effect of entanglement resonance in the anti-parallel
acceleration scenario with the  energy gap   being
$\Omega_{\rm{res}}=\arccos[(2-aL)/2]/(a\sigma^2)$ for any $aL<4$,
which would render the nonlocal correlation term $X$ divergent,
resulting in an infinite concurrence ${\cal{C}}(\rho_{AB})$.
Our numerical integrations
clearly show that all corresponding quantities are finite
and regular, even at point $\Omega=\Omega_{\rm{res}}$ (see
Fig.~(\ref{convsomega1})). This discrepancy is due to the manner of computing the integrals, which  should  be carried out in terms
of the Cauchy principal value  rather than  under a saddle point
approximation as in~\cite{Salton-Man:2015}.  We conclude  that there is
no resonance effect associated with an infinite concurrence
in the anti-parallel acceleration scenario.

\begin{figure}[!htbp]
\centering
\subfloat[$a\sigma=0.01$]{\label{Lmaxvsomega11}\includegraphics[width=0.38\linewidth]{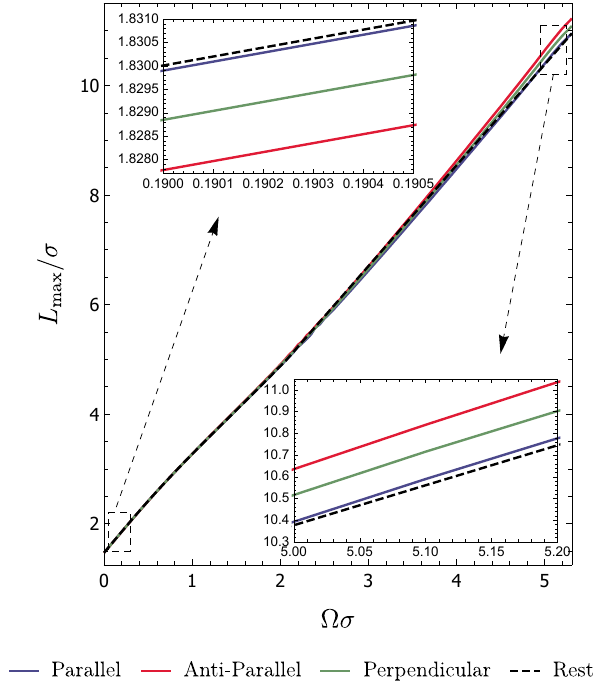}}\qquad
\subfloat[$a\sigma=1.00$]{\label{Lmaxvsomega33}\includegraphics[width=0.38\linewidth]{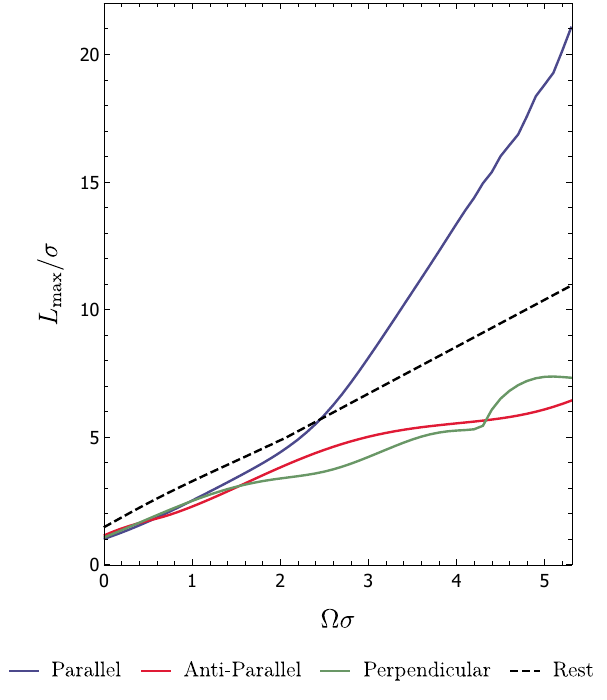}}\qquad
\caption{The maximum harvesting-achievable separation  between two detectors versus the energy gap. Here, we have set $a\sigma=\{0.01\;,1.00\}$ from left-to-right, and the dashed curve denotes the case of detectors at rest.}\label{Lmaxvsomega}
\end{figure}

We now turn to explore the  harvesting-achievable separation range between the detectors in all three  scenarios. For convenience, we introduce a parameter, $L_{\rm{max}}$, to characterize the maximum harvesting-achievable range, beyond which entanglement harvesting cannot occur. As shown in Fig.~(\ref{Lmaxvsomega}), it is easy to see that $L_{\rm{max}}$ is a predominantly increasing function of the energy gap.

For a very small acceleration ($a\sigma\ll1$), the inertial case has the largest
harvesting-achievable range in the small gap regime, followed by parallel, perpendicular, and antiparallel.
This rank ordering tends to converge as the gap increases, and then reverses order for a sufficiently large gap, as shown in
Fig.~(\ref{Lmaxvsomega11})). The anti-parallel acceleration
scenario then becomes the optimum choice that has most ``room" to accessibly
harvest entanglement.

More interestingly, we see in Fig.~(\ref{Lmaxvsomega33}) that for a large, but not too large $a\sigma$, the
parallel acceleration case possesses the
largest $L_{\rm{max}}$, exceeding that   of the inertial detector when $\Omega\sigma \gg\ 1$ and $\Omega>a$.
By contrast, the harvesting-achievable range for the anti-parallel and perpendicular
scenarios cannot exceed that of  the inertial
case.

Summarizing, we find that enhancement of entanglement harvesting happens not only in the anti-parallel
scenario, but also in the parallel and perpendicular
acceleration cases,  in sharp contrast to what was argued previously~\cite{Salton-Man:2015}, namely that there is an enhancement  only compared to  inertial motion only in  the anti-parallel scenario.
Moreover,   the harvesting-achievable  range of acceleration-assisted enhancement  is not universal, holding only for sufficiently large  energy gaps ($\Omega\sigma\gg1$) in the parallel acceleration case, and for sufficiently large  energy gaps with small accelerations ($a\sigma\ll1$)  also in the anti-parallel and perpendicular cases.

\section{conclusion and summary}

In the framework of  the entanglement harvesting protocol, we performed a complete study  on  acceleration assisted entanglement harvesting for two uniformly accelerated detectors in three different scenarios:  parallel, anti-parallel
and mutually perpendicular accelerations. We considered both the amount of entanglement  harvested and the maximal harvesting-achievable separation between  the two detectors. Our numerical evaluation of the integrals involved in the analysis was carried out in terms of the Cauchy principal value,  treating the Wightman function as a distribution rather than using a saddle point approximation as in~\cite{Salton-Man:2015}.

Regarding the amount of harvested entanglement, we find that  the detectors at rest with  small energy gaps $\Omega\sigma < 1$
 harvest comparatively more entanglement than   accelerating ones, suggesting that no acceleration-assisted entanglement harvesting occurs here. In terms of the amount of concurrence harvested, the rank-ordering from largest to smallest is parallel/perpendicular/anti-parallel  for small gaps and separations.  For sufficiently large values of these parameters, this rank-ordering is inverted.
 Furthermore, and quite surprisingly, acceleration can increase
the amount of entanglement harvested for all
three acceleration scenarios once the energy gap is sufficiently large. There exists a peak in
the concurrence at certain  positive energy gap. We find no evidence for  entanglement resonance, as  argued in~\cite{Salton-Man:2015}.

As for the  harvesting-achievable separation range, we find that acceleration hinders entanglement harvesting for detectors with a very small   gap $\Omega\sigma \ll 1$, with inertial detectors (at rest)   possessing a
relatively larger harvesting-achievable range.   However, for a large enough $\Omega\sigma$ and  a small acceleration,
accelerated detectors in all  scenarios possess a comparatively  larger harvesting-achievable range than detectors at rest -- acceleration now assists entanglement. Especially for a not too small acceleration, the parallel  scenario with the gap larger than the acceleration
has the largest harvesting-achievable range amongst all  scenarios, exceeding   that of the inertial case. The
anti-parallel and perpendicular scenarios, however, are unable to overtake the inertial case in terms of their harvesting-achievable range.
These results are in sharp contrast to the previous claim~\cite{Salton-Man:2015}, that
enhancement of the harvesting-achievable range occurs only in the anti-parallel acceleration scenario.

Based on the arguments presented here, a natural question worthy of future study that emerges from
considerations of  the equivalence principle is the extent to which  spacetime curvature  can assist  entanglement harvesting.  Some  investigations in this regard in  AdS spacetime \cite{Zhjl:2019,Ng:2018-2}
and for BTZ black holes \cite{Zhjl:2018,Robbins:2020jca} have shown that curvature can have significant effects on
this process.  However relatively little is known about $(3+1)$ (and higher)-dimensional cases, which merit detailed exploration.  It is also worthwhile to go beyond the massless scalar field case considered here to the massive one and to
other quantum fields.

 \begin{acknowledgments}
 This work was supported in part by the NSFC under Grants No. 11690034, No.12075084 and No.12175062,  the Research Foundation of Education Bureau of Hunan Province, China under Grant No.20B371, and by the Natural Sciences and Engineering Research Council of Canada. RBM would like to thank N. Menicucci and G. Salton for helpful correspondence.
\end{acknowledgments}

\def\ACP{AIP Conf. Proc.}
\def\AIHP{Ann. Inst. Henri. Poincar\'e}
\def\AJP{Amer. J. Phys.}
\def\AM{Ann. Math.}
\def\AP{Ann. Phys. (N.Y.)}
\def\APJ{Astrophys. J.}
\def\ASS{Astrophys. Space Sci.}
\def\ATMP{Adv. Theor. Math, Phys.}
\def\CJP{Can. J. Phys.}
\def\CMP{Commun. Math. Phys.}
\def\CPB{Chin. Phys. B}
\def\CPC{Chin. Phys. C}
\def\CPL{Chin. Phys. Lett.}
\def\CQG{Classcal Quantum Gravity}
\def\CTP{Commun. Theor. Phys.}
\def\EASPS{EAS Publ. Ser.}
\def\EPJC{Eur. Phys.  J. C.}
\def\EPL{Europhys. Lett.}
\def\GRG{Gen. Relativ. Gravit.}
\def\IJGMMP{Int. J. Geom. Methods Mod. Phys.}
\def\IJMPA{Int. J. Mod. Phys. A}
\def\IJMPD{Int. J. Mod. Phys. D}
\def\IJTP{Int. J. Theor. Phys.}
\def\JCAP{J. Cosmol. Astropart. Phys.}
\def\JGP{J. Geom. Phys.}
\def\JETP{J. Exp. Theor. Phys.}
\def\JHEP{J. High Energy Phys.}
\def\JMP{J. Math. Phys. (N.Y.)}
\def\JPA{J. Phys. A}
\def\JPCS{J. Phys. Conf. Ser.}
\def\JPSJ{J. Phys. Soc. Jap.}
\def\LMP{Lett. Math. Phys.}
\def\LNC{Lett. Nuovo Cim.}
\def\MPLA{Mod. Phys. Lett. A}
\def\NPB{Nucl. Phys. B}
\def\PCAM{Proc. Symp. Appl. Math.}
\def\PCPS{Proc. Cambridge Philos. Soc.}
\def\PDU{Phys. Dark Univ.}
\def\PLA{Phys. Lett. A}
\def\PLB{Phys. Lett. B}
\def\PR{Phys. Rev.}
\def\PRA{Phys. Rev. A}
\def\PRD{Phys. Rev. D}
\def\PRE{Phys. Rev. E}
\def\PRL{Phys. Rev. Lett.}
\def\PRX{Phys. Rev. X}
\def\PRSLA{Proc. Roy. Soc. Lond. A}
\def\PTP{Prog. Theor. Phys.}
\def\PRp{Phys. Rept.}
\def\RMP{Rev. Mod. Phys.}
\def\SB{Sci. Bull.}
\def\SPP{Springer Proc. Phys.}
\def\SRTU{Sci. Rep. Tohoku Univ.}
\def\ZPC{Zeit. Phys. Chem.}

\end{document}